\documentclass[amsfonts,amssymb,twocolumn,notitlepage,showpacs,nofootinbib,floatfix]{revtex4-1}
\usepackage{color}
\usepackage{verbatim}
\usepackage{graphicx}
\usepackage{dcolumn}
\usepackage{bm}
\usepackage{amsmath}
\usepackage{bigints}

\begin{document}

\title{Pulsar Timing Perturbations from Galactic\\Gravitational Wave Bursts with Memory}
\author{Dustin R.~Madison}
\email{dmadison@nrao.edu}
\affiliation{National Radio Astronomy Observatory, 520 Edgemont Rd., Charlottesville, VA 22903, USA}
\author{David F.~Chernoff}
\author{James M.~Cordes}
\affiliation{Department of Astronomy, Cornell University, Ithaca, NY 14853, USA}

\date{\today}

\begin{abstract}
\noindent Pulsar timing arrays (PTAs) are used to search for long-wavelength gravitational waves (GWs) by monitoring a set of spin-stable millisecond pulsars. Most theoretical analyses assume that the relevant GW sources are much more distant from Earth than the pulsars comprising the array. Unlike ground- or solar system-based GW detectors, PTAs might well contain embedded GW sources. We derive the PTA response from sources at any distance, with a specific focus on GW bursts with memory (BWMs). We consider supernovae and compact binary mergers as potential Galactic BWM sources and evaluate the signature for an array with pulsars in globular clusters or in the Galactic center. Understanding the response of PTAs to nearby sources of BWM is a step towards investigating other more complex Galactic sources. 
\end{abstract}
\maketitle

\section{Introduction}
Pulsar timing arrays (PTAs) track the rotational phase of
stable millisecond pulsars (MSPs) to search for gravitational
wave (GW) signals with frequencies between approximately $10^{-9}$ and
$10^{-7}$ Hz and strain amplitudes between approximately $10^{-16}$
and $10^{-14}$. Many of the MSPs have been timed for more than ten years with RMS
residuals less than 100~ns \citep{abb+14,abb+16,zhw+14,srl+15,bps+16,ltm+15,tmg+15}. 
The range of GW frequencies and strains made experimentally accessible 
with PTAs encompasses astrophysical sources distinct from those of ground- and
space-based GW detectors and targets hitherto unexplored regions of the
GW parameter space. Supermassive black hole binaries (SMBHBs)
constitute the best known conventional source
population for PTAs but it is possible that exotic, less
well-understood sources like primordial density fluctuations and
cosmic (super)strings will be detected \citep{s13,rws+15,sbs12,ltm+15,abb+16}.
Such a detection would be both revolutionary and important.

Ground- and space-based GW detectors are several kilometers and
several million kilometers in size, respectively, while the pulsars
involved in PTA efforts are located kiloparsecs from Earth. The ratio
of the size of the detector to the GW source distance is an intrinsic
parameter affecting the nature of the waveform incident on the
detector. For ground- and space-based detectors, the ratio is
always much less than one, even for GW sources within our own Galaxy. For
PTAs, on the other hand, the ratio is source dependent. It is very
small for SMBHBs located hundreds of megaparsecs away and in
such a case the commonly used plane-parallel treatment for the GWs
is completely adequate. But for sources within the Galaxy, the
plane wave limit is inadequate \citep{s78,ksg+99,lb01,df11,krp12}.

When a source ejects particles and/or radiation in an asymmetric fashion,
one consequence is a permanent change in the distance between
freely-falling test masses. Linear and nonlinear terms in the Einstein
equations couple components of the varying stress energy tensor to
give a net change in the metric. This is memory, an effect that
generically accompanies the violent processes that lead to GW
emission.  In this paper, we calculate the PTA response to a GW burst
with memory (BWM).  Explosions and inspirals provide astrophysical
situations where such an effect is to be expected.  The characteristic
timescale for the memory to reach its final value is the timescale for
the explosion or inspiral \citep{bt87,c91,t92,f09,pr11}.

In bursting sources such as core-collapse supernovae, memory grows to
its final value on the dynamical timescale of the newly-formed neutron
star. The situation for inspirals is
less clearcut since the process is a long event punctuated by the
final merger.  Presumably the magnitude of
the instantaneous contributions to the final metric change correlates
with the instantaneous amplitude of GW emission from the source.
Current post-Newtonian estimates suggest that for merging binaries,
the accumulated memory effect increases gradually through the long
early stages of inspiral and then rapidly in the final chirp and
plunge. The effect need not grow monotonically since the metric
change is a signed quantity. The final value has not yet been
calculated for binary mergers
by numerical relativity, the only reliable
description of the most important phases, the plunge and coalesence.
In any case, we will treat both explosions and inspirals as
instantaneous events and the magnitude of the final memory as a
critical parameter that must be estimated.

Several works have detailed the PTA response to BWMs in the plane wave
limit \citep{s09,pbp10,vl10,cj12,mcc14} and derived upper limits on
the amplitude and the rate of events
\citep{abb+15,whc+15,mzh+16}. Today's constraints currently lie above
theoretically predicted levels due to the SMBHB population. Since many
GW sources can produce memory effects, BWM searches serve
to discover and/or limit a wide class of phenomena
\citep{cbv+14}. Furthermore, apart from the details of any particular
source, the mere detection of memory would be important as a physical
test of deep concepts in general relativity. \citet{s17} describes
memory as a consequence of infinite symmetries and conserved
quantities in general relativity, a cornerstone concept for
understanding the infrared sector of the theory.

Since BWMs likely accompany a wide array of GW events and have
particularly simple time-domain behavior, they are useful prototypical
signals for exploring near-field GW effects in a PTA's response. In
Section~II of this paper, we derive the pulsar timing perturbation
produced by a BWM, compare our result to the the plane parallel
approximation, and consider feasible astrophysical sources for
Galactic BWMs---supernovae and inspiral mergers of compact
binaries. In Section~III, we investigate how Galactic BWMs influence
the observed properties of pulsars and how the observable timing
perturbation produced by a Galactic BWM is modified by the model
fitting used in pulsar timing analyses. We also discuss how timing pulsars in
dense stellar environments such as globular clusters and the galactic
center may be particularly advantageous for detecting Galactic
BWMs. Finally, in Section~IV, we discuss our work and offer some
concluding remarks.

\section{Timing Perturbation from a BWM}

Consider a GW burst from a source located at the coordinate origin.
The persistent change in the metric at a
field point a distance $r$ from the origin \citep{bt87} is
\begin{eqnarray}
\label{braginskii_thorne}
h_{ij}^{\rm mem}=\delta\sum_{A=1}^N\frac{4GE_A}{c^6r}\left[\frac{v_A^iv_A^j}{1-\left(\displaystyle\frac{v_A}{c}\right)\cos{\theta_A}}\right]^{\rm TT}, 
\end{eqnarray}
where the sum is over gravitationally independent groups of particles with constant 4-momenta labeled by ``A''. The $\delta$ indicates that the sum before the burst is to be subtracted from the sum after the burst. The number of terms in the summation can vary, e.g. a single particle can explode into many particles. The energy of the $A$th component of the system is $E_A$ and $v_A$ is its velocity. The angle between $v_A$ and the direction to the field point is $\theta_A$. For particles with invariant mass $M_A$, the energy is $E_A=M_Ac^2 \left( 1-(v_A/c)^2 \right)^{-1/2}$. For massless particles, $v_A=c$ and $E_A$ is the energy. 

Assume a mass $M$ at rest at the origin explodes at
time $t=t_{burst}$ giving two equal masses that travel along the positive and negative $z$ axes at equal speed $v$. 
The sum in Equation~\ref{braginskii_thorne} vanishes before the burst. The initial energy of the system, $E=Mc^2$, is equally split between the two outgoing particles and energy conservation requires that the invariant mass of each outgoing particle be $(M/2)\left( 1-(v/c)^2 \right)^{1/2}$. The metric at point ${\bf r}$ at time $t$ is
\begin{eqnarray}
\label{specificExampleMemory}
h_{ij}^{\rm mem}(t,{\bf r})&=&\frac{4GEv^2\Theta(t_{\rm ret}(t,{\bf r})-t_{burst})}{c^6r\left[1-\left(\displaystyle\frac{v}{c}\right)^2\cos^2{\theta}\right]}\nonumber\\
&  &\Lambda_{ij,kl}\hat{z}^k\hat{z}^l,
\end{eqnarray}
where $\Theta$ is the Heaviside step function and
the retarded time $t_{\rm ret}(t,{\bf r})=t - r/c$. The angle between the z-axis and the field direction is $\cos{\theta}=\hat{{\bf z}}\cdot\hat{{\bf r}}$ (we use $\hat{\bf x}$, $\hat{\bf y}$, and $\hat{\bf z}$ as the unit basis of the center-of-mass coordinate system) and $\Lambda_{ij,kl}=P_{ik}P_{jl}-(1/2)P_{ij}P_{kl}$ enforces transverse-traceless gauge at the field point where the projection operator $P_{ij}=\delta_{ij}-\hat{r}_i\hat{r}_j$. 

Suppose the Earth is located at position ${\bf d}_E$, a pulsar is located at position ${\bf d}_P$, and both are at rest relative to the BWM source's center of mass (with distances $d_E=|{\bf d}_E|$, $d_P=|{\bf d}_P|$ respectively)\footnote[1]{The Earth, of course, orbits the solar system barycenter (SSB) which will itself be moving at an approximately fixed velocity relative to any BWM source. Converting temporal measurements from observatories on the Earth's surface to the inertial SSB frame is common pulsar timing practice so we will continue to refer to ${\bf d}_E$ as the location of the Earth. Relative motion between the Earth, BWM source, and pulsar can be treated using the methods we develop, but to simplify our calculations and reduce the size of the relevant parameter space, we opt to ignore it.}. Let the distance from pulsar to Earth be $d$.

The arrival time perturbation of pulsar photons detected at Earth is
\begin{eqnarray}
\label{timingFormalism}
\Delta(t)=\frac{1}{2}\int_{t-d/c}^t\hat{k}^i\hat{k}^jh^{\rm mem}_{ij}\left(t',{\bf r}(t')\right)dt'.
\end{eqnarray}
The integral is over the unperturbed path of the photons traveling from the
pulsar to the Earth and arriving at time $t$. The endpoints satisfy ${\bf r}(t)={\bf d}_E$ and ${\bf
  r}(t-d/c)={\bf d}_P$.  In addition to the first order, perturbative
treatment of the photon's motion in spacetime we make two basic
assumptions. First, the light travels between pulsar and Earth
exclusively in the radiative zone of the BWM source. That zone begins at
least a few gravitational wavelengths from the source. This is
generally an easy condition
to satisfy for a particular pulsar-Earth path because the zone is
small compared to the characteristic Galactic scale. This is
true even for sources
that lie within a PTA detector. Second,
Equation~\ref{specificExampleMemory} uses $\theta(t_{ret}-t_{burst})$ for
the BWM waveform.  The characteristic time it takes the memory component
to grow from zero to its final value at a field point depends upon the
physical process associated with the burst. It is assumed to be short. The time for an explosive
event is a dynamical time, e.g. $\sim 10^{-2}$ s for neutron star
collapse and bounce.  For merging binaries most of the memory effect
probably accumulates during the plunge and coalesence so the time
scale is the light travel time across the post-merger
object---kilometers for stellar mass black holes and less than an AU
for supermassive black holes. All these timescales are short compared
to inverse frequencies over which PTAs are sensitive, $10^7-10^9$ s.

\subsection{Evolution of the Timing Perturbation}

The non-relativistic limit of Equation~\ref{specificExampleMemory} is
\begin{eqnarray}
\label{nonRel}
h_{ij}^{\rm mem}(t,{\bf r})&=&\frac{4GEv^2}{c^6r}\Theta(t_{\rm ret}(t,{\bf r})-t_{burst}) \nonumber \\
& & \Lambda_{ij,kl}\hat{z}^k\hat{z}^l
\left[1+{\cal O}\left(\left(\frac{v}{c}\right)^2\right)\right] .
\end{eqnarray}
We now complete the description of the geometry.  Let $\hat{\bf k}$ be
the unit vector pointing from the pulsar to the Earth.  This is the
direction traveled by the photon moving toward the Earth bound
observer. Let $\hat{\bf b}$ be the unit vector from the source to the
point of closest approach along the photon path. The impact parameter
is $b$ and the path is perpendicular ($\hat{\bf k}\cdot\hat{\bf b}=0$)
at that point. Finally, let the angle between the source-to-Earth
direction and the photon path be $\cos \beta = {\hat {\bf k}} \cdot
{\hat {\bf d}}_{\bf E}$. In this geometry the impact parameter $b=d_E
\sin \beta$ and the distance from the point of closest approach to
Earth is $d_E \cos \beta$.

Combining Equations~\ref{timingFormalism} and \ref{nonRel} yields an
explicit quadrature for the timing perturbation. We convert the
integral over time $\int^t_{t-d/c} dt' \cdots $ to the form
$\int_{u_-}^{u_+} (dt'/du) du \cdots$ where $u=(ct'- ct + d_E \cos
\beta)/b$ and the endpoints are
$u_+=u(t)=d_E \cos \beta/b = \cot \beta$ and $u_-=u(t-d/c)=u_+ - d/b$. We
find
\begin{eqnarray}
\label{nonRelativistic}
\Delta(t)&=&\frac{GMv^2}{c^5 }\int_{u_-}^{u_+}\frac{\alpha_0+\alpha_1u+\alpha_2u^2}{(1+u^2)^{5/2}}\nonumber\\& &\times~\Theta\left[t-\frac{d_E}{c}\cos{\beta}-\frac{b}{c}f(u) - t_{burst}\right]du,
\end{eqnarray}
where
\begin{eqnarray}
\alpha_0&=&2\hat{k}_z^2+\hat{b}_z^2-1,\\
\alpha_1&=&-2\hat{k}_z\hat{b}_z,\\
\alpha_2&=&\hat{k}_z^2+2\hat{b}_z^2-1,\\
f(u)&=&\sqrt{1+u^2}-u.
\end{eqnarray}
Although the step function has been transformed to be an explicit
function of $t$, $t_{burst}$, $u$ and geometric quantities, it is equivalent to the
simple form $\Theta(t_{ret}-t_{burst})$ which is zero for field points
with $t_{ret}(t,{\bf r}) < t_{burst}$ or, equivalently, $t < r/c +
t_{burst}$.  Non-zero contributions to $\Delta(t)$ begin when the BWM
arrives at the Earth, so that $t=d_E/c+t_{burst}$. Although the BWM
may intersect the path prior to this instant, the perturbed photons
take time to travel to Earth and arrive there later. Without loss of
generality, choose $t_{burst}=-d_E/c$ so that $\Delta(t)=0$ for $t<0$.
This choice is convenient because all timing effects begin at
$t=0$. The timing perturbation is initially sourced by metric
perturbations at spacetime points near ${\bf r}={\bf d}_E$ and
$t=0$. The effect accumulates until the whole journey of the photons
occurs within the expanding BWM wavefront.

The integration gives
\begin{eqnarray}
\label{solution}
\Delta(t)=\frac{\Delta_z}{6}\frac{\alpha_0u(3+2u^2)-\alpha_1+\alpha_2u^3}{(1+u^2)^{3/2}}\biggr|^{u_+}_{u_L(t)},
\end{eqnarray}
where  
\begin{eqnarray}
\label{delta0}
\Delta_z&=&\frac{2GMv^2}{c^5},\nonumber\\
&\approx&0.1~{\rm ns}~\left(\frac{M}{M_\odot}\right)\left(\frac{v}{10^3~{\rm km}~{\rm s}^{-1}}\right)^2,
\end{eqnarray}
and the condition for the step function to be unity is $u_L(t) < u < u_+$
with
\begin{eqnarray}
u_L(t)={\rm max}\left[u_-,f^{-1}\left(f(u_+)+\frac{ct}{b}\right)\right].
\end{eqnarray}
Here $f^{-1}$ is the function inverse of $f$.
For bursts of massless particles, $\Delta_z=2GE/c^5$ (the factor of two in our definition of $\Delta_z$ is convenient for derivations in Appendix A). At time
 \begin{eqnarray}
 \label{finalTime}
 t_F=\frac{b}{c}\left[f(u_-)-f(u_+)\right],
 \end{eqnarray}
 the quantity $u_L(t_F) \to u_-$ and the timing perturbation reaches its asymptotic value $\Delta_F$, $\Delta(t>t_F)=\Delta_F$.
 
If one fixes the location of the Earth and the direction to the pulsar
then the distance between Earth and pulsar influences $\Delta(t)$
through $u_L(t)$ and specifically via the dependence of $u_-$ on $d$. Consider two such pulsars
with distances $d_1 < d_2$. The timing perturbations for the nearer
one saturates first so that $t_{F1} < t_{F2}$ but $\Delta(t)$ is {\it the
  same} for both pulsars for times $t< t_{F1}$. In other words, there
is a common asymptotic form for $\Delta(t)$ which is truncated at
different final times as determined by the distance $d$.  The asymptotic
change $\Delta_F$ depends on $d$.

We can compare Equation~\ref{solution} with plane-wave descriptions commonly used in the literature on BWMs and pulsar timing. In the plane wave limit, the timing perturbation of a BWM of amplitude $h_B$ striking the Earth at time $t_0$ is 
\begin{eqnarray}
\label{planeWaveLimit}
\Delta_P(t)& = & \frac{h_B}{2}\cos{2\varphi}(1+\cos{\beta})\nonumber\\
 & &\left[(t-t_0)\Theta(t-t_0)-(t-t_1)\Theta(t-t_1)\right],
\end{eqnarray}
where $t_1 = t_0+(d/c)(1-\cos{\beta})$ and $\varphi$ is the angle between the wave's principal polarization vector and the projection of the pulsar line of sight onto the plane normal to the wave propagation direction \citep{pbp10,vl10,cj12,mcc14,abb+15,whc+15}. The subscript ``$P$" in $\Delta_P$ denotes the plane-wave limit. The expression for $\Delta(t)$ in Equation \ref{solution} reduces to $\Delta_P(t)$ in the appropriate limiting case (see Appendix~A where we relate
$h_B$ and $\varphi$ to the variables in Equation~\ref{solution}). Equation~\ref{planeWaveLimit} implies that $\Delta_P$ grows to a final value
\begin{eqnarray}
\Delta_{P,F} = \frac{h_Bd}{2c}\cos{2\varphi}(1+\cos{\beta})(1-\cos{\beta}),
\end{eqnarray}
after a time $t_{P,F}=t_1-t_0=(d/c)(1-\cos{\beta})$ has elapsed.

\begin{figure}
\begin{center}
\includegraphics[scale=.45]{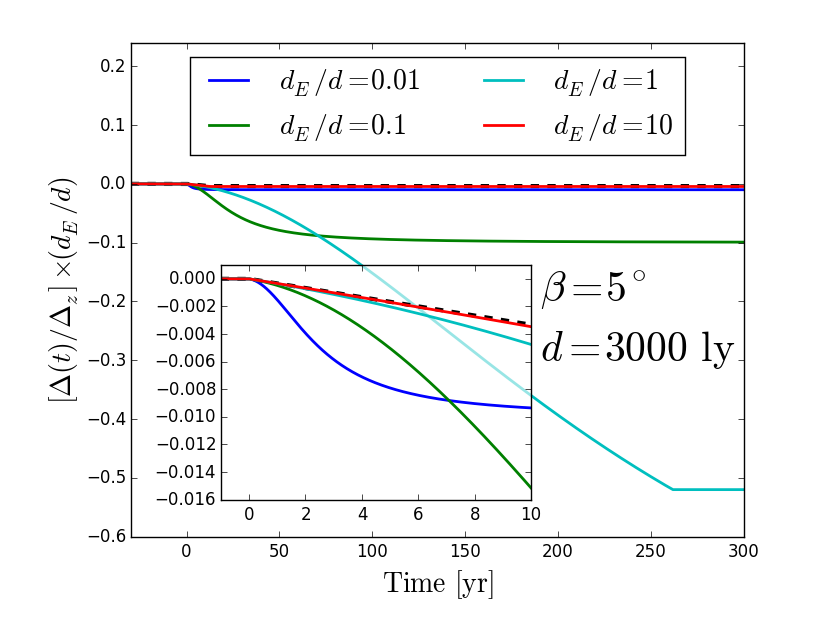}
\includegraphics[scale=.45]{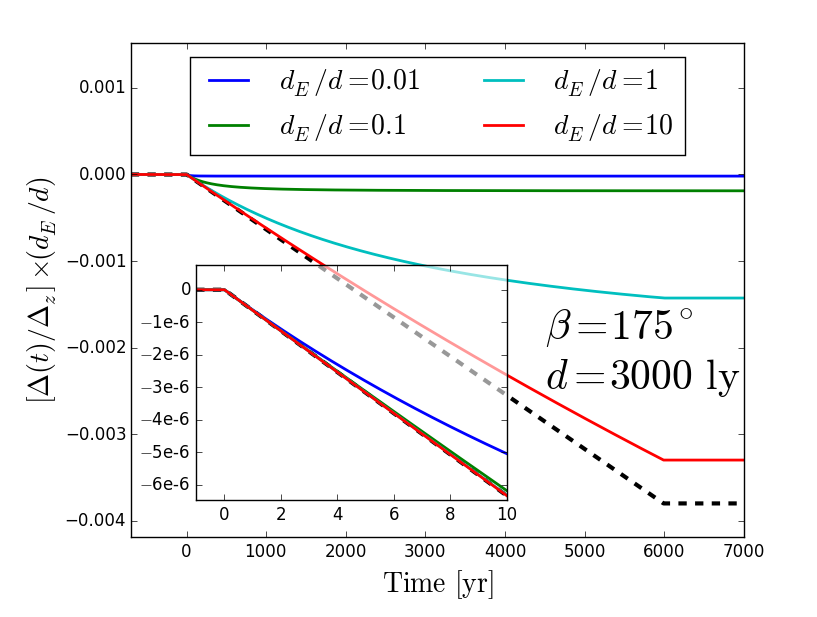}
\caption{The full evolution of $\Delta(t)$ for a variety of geometric configurations of Earth, pulsar, and BWM source. The inset in each panel is a close-up of the first 10 years of evolution after the memory wavefront passes through the solar system. We plot $\Delta(t)$ multiplied by $d_E/d$ (holding $d$ fixed) to counteract the $d_E^{-1}$ scaling of the plane wave limit. Without near-field effects, all curves would follow the dashed black curve.}
\end{center}
\end{figure}

In Figure~1, we compare the evolution of $\Delta(t)$ and $\Delta_P$
for several configurations of the Earth, pulsar, and BWM source. We
confine the Earth-pulsar system to the $x$-$y$ plane, where
$\theta=\pi/2$. In this plane, Equation~\ref{specificExampleMemory}
reduces to Equation \ref{nonRel}, i.e. any relativistic modifications
to the BWM vanish.  We fix $d=3000$~lyr, consider four values of $d_E$
(the four curves in each panel), and angles $\beta=5^\circ$ (top
panel) and $\beta=175^\circ$ (bottom panel). These choices facilitate
comparison between $\Delta$ and $\Delta_P$. The latter has a
particularly simple scaling $\Delta_P(t) \propto h_B \propto d_E^{-1}$
since $\varphi$ is constant for Earth and pulsar in the $x$-$y$ plane.

Figure 1 shows $\Delta_P(t)$ by the dashed black curve and $\Delta(t)$
by the colored lines. We have multiplied $\Delta$ and $\Delta_P$ by
$d_E/d$ so that all curves would match the dashed black curve if the
plane-wave description were accurate. Deviations from the dashed black
curve are due to near-field effects such as wavefront curvature and
non-negligible variations in the amplitude of the BWM along the photon
trajectory. These deviations highlight the new results of this
calculation. The insets in each panel of Figure~1 magnify the 10 years
immediately following the BWM's arrival.  The differences are most
apparent for $\beta$ near $0$ and $\pi$ when the impact parameter is small and motivate the two choices. In
Table~I, we list $t_F$, $t_{P,F}$, $\Delta_F/\Delta_z$, and
$\Delta_{P,F}/\Delta_z$ for all curves in Figure~1.

Since the source is axisymmetric in the $x$-$y$ plane, the GW amplitude in the plane varies solely with the distance from the source rather than its orientation or structure. Any axisymmetric BWM source effectively radiates isotropically and will produce timing perturbations identical to those depicted in Figure~1, though with an amplitude other than $\Delta_z$. We will see this later when we consider supernovae and inspiral mergers.

\begin{table}[]
\begin{equation}
\begin{array}{|c|c|c|c|c|c|}
\hline
~\beta~&~d_E~&~t_F~&~t_{P,F}&~\Delta_F/\Delta_z~&~\Delta_{P,F}/\Delta_z~\\
(\rm{deg})&(\rm{lyr})&(\rm{yr})&(\rm{yr})&&\\
\hline\hline
5&3\times10^1&5940.1&11.4&-0.99810&-3.8\times10^{-1}\\
5&3\times10^2&5401.3&11.4&-0.99807&-3.8\times10^{-2}\\
5&3\times10^3&261.7&11.4&-0.519907&-3.8\times10^{-3}\\
5&3\times10^4&12.7&11.4&-0.000445&-3.8\times10^{-4}\\
\hline
175&3\times10^1&5999.9&5988.6&-0.001902&-3.8\times10^{-1}\\
175&3\times10^2&5999.0&5988.6&-0.001887&-3.8\times10^{-2}\\
175&3\times10^3&5994.3&5988.6&-0.001427&-3.8\times10^{-3}\\
175&3\times10^4&5989.6&5988.6&-0.000330&-3.8\times10^{-4}\\
\hline
\end{array}\nonumber
\end{equation}
\caption{Numbers descriptive of Figure~1. The top four rows correspond to the upper panel of Figure~1 and the bottom four rows correspond to the lower panel of Figure~1.}
\end{table}

\begin{figure}
\begin{center}
\includegraphics[scale=.45]{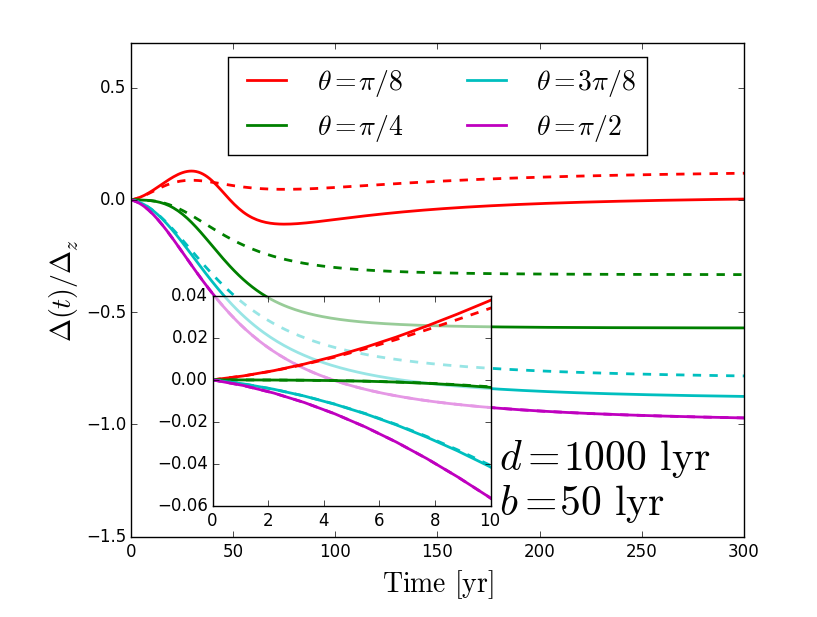}
\includegraphics[scale=.45]{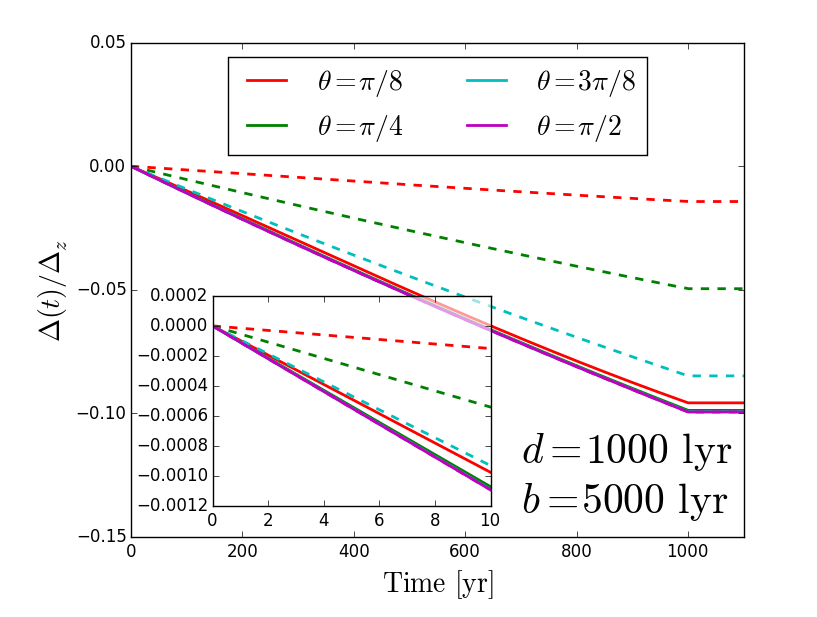}
\caption{The timing perturbation for a symmetric burst of relativistic particles (solid curves) and for non-relativistic particles (dashed curves) for several geometric configurations of Earth, pulsar, and BWM source. The line connecting the Earth and pulsar is kept perpendicular to and symmetric about the exploding source's symmetry axis and the polar angle of the vector $\hat{b}$ is varied between four values of $\theta$.} 
\end{center}
\end{figure}

\subsection{Relativistic Motions}
Relativistic motions modify the timing perturbation when $\theta\neq\pi/2$. In this case, the integral describing the timing perturbation is
\begin{eqnarray}
\label{relativistic}
\displaystyle\frac{\Delta(t)}{\Delta_z}&=&\frac{1}{2}\bigintss_{u_L(t)}^{u_+}\frac{\displaystyle(\alpha_0+\alpha_1u+\alpha_2u^2)du}{\displaystyle(1+u^2)^{5/2}\left[1-\left(\frac{v}{c}\right)^2\frac{(u\hat{k}_z+\hat{b}_z)^2}{(1+u^2)}\right]}.\nonumber\\
\end{eqnarray}
We compute this integral numerically. In Figure~2, we demonstrate the differences between the relativistic and non-relativistic solutions. We fix $\hat{\bf k}=\hat{\bf y}$ with the Earth and pulsar situated symmetrically about the $x$-$z$ plane. For the top panel, we set $d=1000$ lyr and $b=50$ lyr; for all values of $\theta$, $\beta\approx6^\circ$. For the bottom panel, we kept $d=1000$ lyr but set $b=5000$ lyr; for all values of $\theta$, $\beta\approx84^\circ$. We vary $\hat{\bf b}=\sin{\theta}\hat{\bf x}+\cos{\theta}\hat{\bf z}$. The solid lines show the relativistic result computed using Equation~\ref{relativistic} (with $v=c$) while the dashed lines represent the non-relativistic results from Equation~\ref{solution}. As with Figure~1, the inset in each panel shows only the ten years following the BWM passing through the solar system.

The top panel of Figure~2 demonstrates that in the 10 years following the BWM, the relativistic solution is almost indistinguishable from the non-relativistic solution, but within approximately 50 yr, the differences become significant. For small values of $\theta$ and $b$ (the red curves in the top panel in Figure~2), we see that non-monotonic evolution of the timing perturbation is possible. The bottom panel of Figure~2 demonstrates that for large values of $b$, relativistic effects eliminate variations in the magnitude of the timing perturbation with variations in $\theta$, i.e. the relativistic burst, when viewed from a great distance, is a more nearly isotropic radiator of memory than the non-relativistic burst.

\subsection{Supernovae}

\begin{figure}
\begin{center}
\includegraphics[scale=.45]{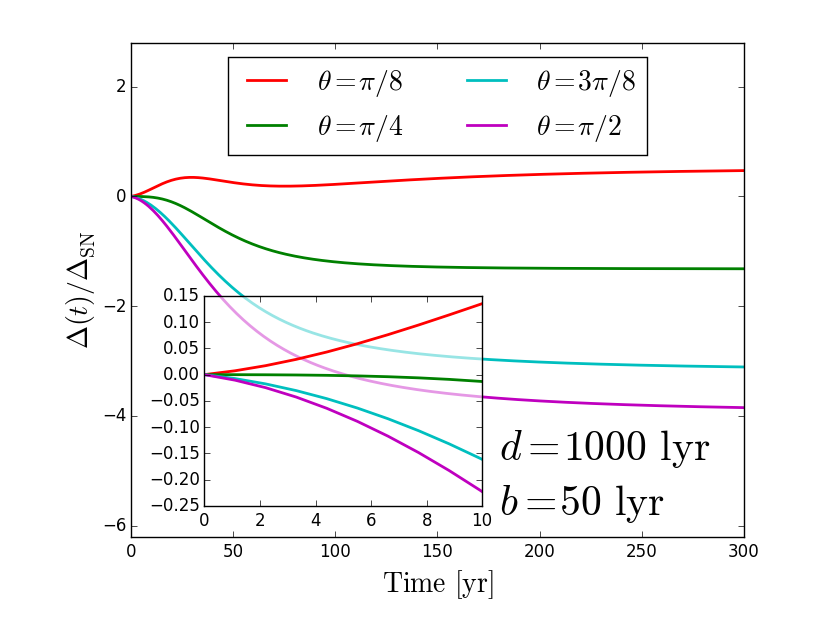}
\includegraphics[scale=.45]{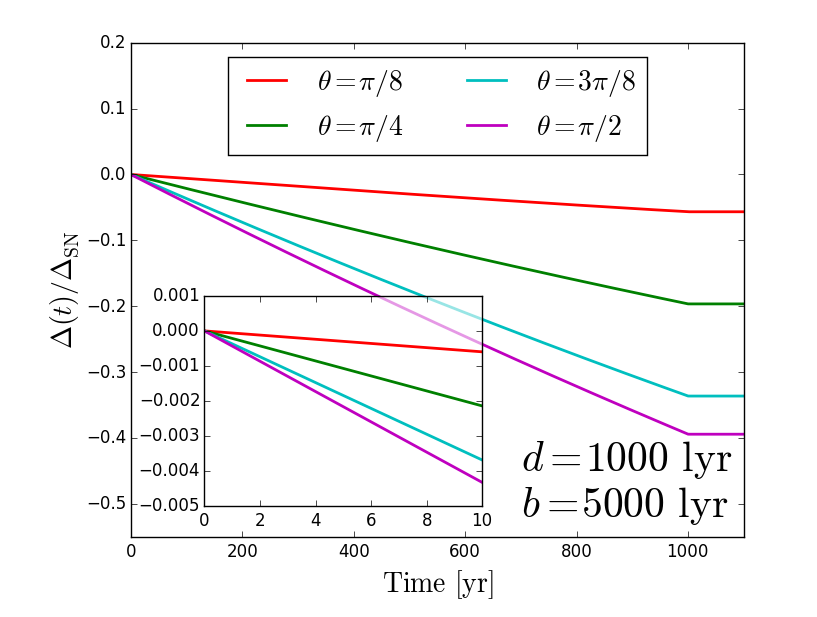}
\caption{The timing perturbation produced by the explosion of relativistic particles from a supernova. We have simplistically modeled the distribution of released energy as proportional to the spherical harmonic $Y_{20}(\theta,\phi)$. The configuration of Earth, pulsar, and BWM source is the same for each curve as it was in Figure~2.}
\end{center}
\end{figure}

Asymmetry in a core-collapse supernova imparts a natal kick to a stellar-mass compact remnant, often resulting in velocities of hundreds of kilometers per second, quite like the idealized symmetric exploding mass we investigated above \cite{cc98}. They are also observed to occur approximately once per century per galaxy. As such, supernovae are potential sources of Galactic BWMs. But the kinetic energy of the escaping compact remnant is only a fraction of the total energy released by the supernova. The rest of the energy is jettisoned as relativistic particles---neutrinos, photons, and GWs. 

We describe the energy distribution of relativistic particles with spherical harmonics:
\begin{eqnarray}
\frac{dE}{d\Omega}=\sum_{l,m}b_{lm}Y_{lm}(\theta,\phi).
\end{eqnarray}
The $l=0$ moment determines the total energy of the relativistic particles: $E=\sqrt{4\pi}b_{00}$. The $l=1$ moments determine their linear momentum, which counters the momentum of the ejected compact remnant. If we assume the remnant is kicked along the $z$ axis, its momentum is $cp_z=-\sqrt{4\pi/3}b_{10}$. 

The BWM produced by these particles at a field point in direction $\bf{\hat{n}}$ is 
\begin{eqnarray}
\label{shellMetric}
h^{\rm mem}_{ij,B}=\frac{4G\Theta(t_{\rm ret})}{c^4r}\Lambda_{ij,kl}\int d\Omega \frac{dE}{d\Omega}\frac{\hat{\zeta}^k\hat{\zeta}^l}{1-\hat{\bf n}\cdot\hat{\bf \zeta}}.
\end{eqnarray} 
We assume the velocity of outgoing particles is ${\bf v}=c\hat{\zeta}$. In Appendix B, we work out the form of the BWM produced by the flash of relativistic particles described by Equation~\ref{shellMetric}. Moments with $l\geq2$ contribute to the memory. We then compute the timing perturbation when the only nonzero moment with $l\geq2$ is $b_{20}$ (making the source symmetric about the $x$-$y$ plane as in the exploding mass example treated above). This is not an altogether unreasonable simplification as rotation in a progenitor star will generate perturbations in the star's density profile proportional to $Y_{20}(\theta,\phi)$ \citep{fka+15}. 

In Figure~3, we adopt the same geometric configuration of source, Earth, and pulsar as used to produce Figure~2. The results for SN in Figure~3 resemble the non-relativistic curves shown in Figure~2; the amplitude of memory radiated by the non-relativistic symmetric exploding mass and our SN model vary in $\theta$ in the same fashion. We have scaled the curves in Figure~3 by $\Delta_{SN}=Gb_{20}/c^5$ rather than $\Delta_z$ If we say $b_{20}=\epsilon_2b_{00}$, 
\begin{eqnarray}
\Delta_{SN}\approx2.7~{\rm ns}~\left(\frac{\epsilon_2}{0.01}\right)\left(\frac{b_{00}}{10^{53}~{\rm ergs}}\right).
\end{eqnarray}

For non-relativistic velocities, the BWM produced by a single mass being suddenly kicked away from the origin is identical to the non-relativistic limit of Equation~\ref{specificExampleMemory}. 
Writing $|b_{10}|=\epsilon_1b_{00}$, the velocity of the kicked remnant (of mass $M$) will be
\begin{eqnarray}
|v_z|\approx341~{\rm km}~{\rm s}^{-1}\left(\frac{\epsilon_1}{0.01}\right)\left(\frac{M}{M_\odot}\right)^{-1}\left(\frac{b_{00}}{10^{53}~{\rm ergs}}\right),\nonumber\\
\end{eqnarray}
a safely non-relativistic value. Plugging this velocity into Equation~\ref{delta0} and noting the similarities between Figures~2 and 6, it is clear that the kicked remnant contributes substantially less to the magnitude of the memory than the relativistic particles produced by the SN.

\subsection{Inspiral Mergers}
Inspiraling compact binaries, since they involve stellar masses and relativistic velocities, are anticipated to be sources of strong BWMs. Since the memory of such systems grows over their entire inspiral history, the analysis we have done in this paper involving instantaneous bursts is not entirely appropriate for treating this problem. Nonetheless, the memory is predicted to grow primarily during the final stages of the merger on a timescale approximately equal to a few times the light-crossing time of the post-merger event horizon, effectively instantaneously given the weekly or monthly observing cadence typical of pulsar timing \cite{f09}. 

For the symmetric exploding mass, Equation~\ref{specificExampleMemory} shows that the memory amplitude $h^{\rm mem}_{ij}\propto c\Delta_z/r$ times a function of order unity describing the geometry and radiation pattern of the source---a beam function. In Appendix B, we show that for a supernova, $h^{\rm mem}_{ij}\propto c\Delta_{\rm SN}/r$ times a beam function. Similarly, the amplitude of the memory produced by inspiral mergers is $h_m=\xi G\mu/c^2r$ times a beam function where $\mu$ is the reduced mass of the system and $\xi\approx0.04$ \cite{f09,abb+15}. In all of these cases, the beam function is azimuthally symmetric. So, similar to our definitions for $\Delta_z$ and $\Delta_{\rm SN}$, define
\begin{eqnarray}
\label{inspiral}
\Delta_m=\frac{\xi G\mu}{c^3}\approx2~\mu{\rm s}~\left(\frac{\xi}{0.04}\right)\left(\frac{\mu}{10~M_\odot}\right).
\end{eqnarray}
We do not work out the timing perturbation caused by a local inspiral merger in detail, but as the three sources we have considered all have azimuthally symmetric beam functions, Figure 1 scaled by $\Delta_m$ rather than $\Delta_z$ or $\Delta_{\rm SN}$ accurately describes the timing perturbation that is caused by a merging binary when the Earth and pulsar are confined to the binary's orbital plane. 

LIGO recently detected the merger of approximately 30 $M_\odot$ binary black holes \citep{aaa+16}. Though such events are rare in our Galaxy, given the scale of $\Delta_m$, such a merger (in our Galaxy) could produce a timing perturbation exceeding several microseconds over 10 to 20 years, a feasibly detectable effect since many MSPs can be timed with a residual RMS less than several hundred nanoseconds over decadal time scales.

\subsection{Final Timing Perturbation Compared to\\Limiting Plane Wave Value}

We have thus far investigated the evolution of the timing perturbation caused by a local BWM. Here we compare the final timing perturbation $\Delta_F$ to the value anticipated using the plane wave approximation, $\Delta_{P,F}$. We define two dimensionless parameters with which to specify the relative locations of the Earth, pulsar, and BWM source. First, $\psi\equiv b/d$, the impact parameter divided by the distance between the Earth and pulsar. Second, $\chi$, the distance from the Earth-pulsar midplane divided by $d$ with positive $\chi$ on Earth's side of the midplane. In terms of these parameters, the pulsar is located at $\psi=0$ and $\chi=-1/2$ while the Earth is located at $\psi=0$ and $\chi=+1/2$. We place the BWM source at a variety of values of $\chi$ and $\psi$ and compute $\Delta_F$ and $\Delta_{P,F}$. Their ratio is depicted in Figure~4.

\begin{figure}
\begin{center}
\includegraphics[scale=.45]{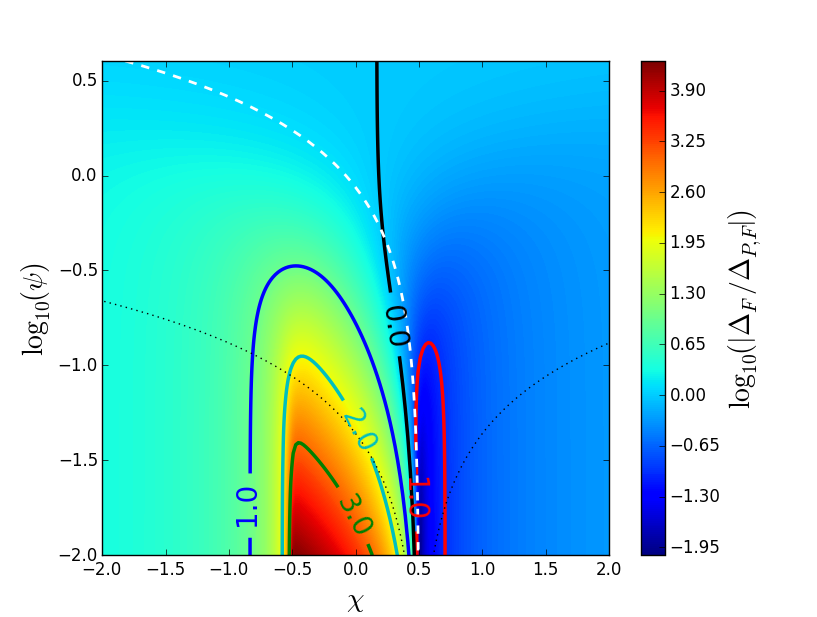}
\caption{The magnitude of the final timing perturbation, $\Delta_F$, divided by $\Delta_{P,F}$, the magnitude anticipated by commonly used plane-wave descriptions of BWMs. The Earth and pulsar are positioned at $\psi = 0$ and $\chi=+1/2$ and $-1/2$, respectively. The memory source is centered on the variable position ($\chi$,$\psi$) and the Earth-pulsar system has been confined to the memory source's symmetry plane. The dotted black curves are rays of constant $\beta$ emanating from the Earth's position (from left to right, they are $5^\circ$ and $175^\circ$). The dashed white curve is a ray of $\beta=60^\circ$. We discuss these rays in the text.}
\end{center}
\end{figure}

The black contour in Figure~4 shows where $\Delta_{P,F}$ and $\Delta_F$ are equal. As $\psi\rightarrow\infty$, the black contour continues to extend vertically upward and asymptotically approaches $\chi=0$. Everywhere right of the black contour the plane wave approximation is an overestimate for $\Delta_F$, and everywhere left of the black contour, it is an underestimate.   

In a small region where $1/2\lesssim\chi\lesssim3/4$ and $\psi \lesssim 0.1$ (inside the red contour), the plane wave approximation overestimates the final timing perturbation by more than a factor of 10. For $-1/2\lesssim\chi<+1/2$, and $\psi\lesssim10^{-0.5}$ (for Earth-pulsar separations of 1 kpc, this corresponds to impact parameters less than approximately 300 pc), and more specifically below the dashed white curve depicting a ray of constant $\beta=60^\circ$, $\Delta_F$ exceeds $\Delta_{P,F}$ by factors of more than 10, 100, and even 1000 (blue, cyan, and green contours, respectively). The impact parameter is $d_E\cos\beta$, or $d_E/2$ along the dashed white curve where $\beta=60^\circ$. This particular value of $\beta$ demarcates a qualitative transition in the behavior in time of the BWM-induced timing perturbation; we will discuss this more in the next section.

The dotted black curve moving up and to the left corresponds to $\beta=5^\circ$. The dotted black curve moving up and to the right corresponds to $\beta=175^\circ$. One can see from Figure~4 that in the region between the Earth and pulsar, the ray of $\beta=5^\circ$ moves to higher values of $|\Delta_F/\Delta_{P,F}|$ as $d_E$ is increased. Said otherwise, in this region, the plane-wave approximation, counterintuitively, becomes increasingly incorrect as the distance from the Earth to the source is increased. 

The curves in Figure~1 are all consistent with what is shown in Figure~4. When $\beta=175^\circ$ (corresponding to the bottom panel of Figure~1), the BWM source is always right of the solid black contour in Figure~4 and the magnitude of the timing perturbation is always smaller in magnitude than what is anticipated from the plane-wave approximation, including when it saturates at $\Delta_F$. When $\beta=5^\circ$ (the top panel of Figure~1), the BWM source is always left of the black contour in Figure~4 and the magnitude of the final timing perturbation is always greater than what is anticipated from the plane-wave approximation. 

When discussing the centermost part of Figure~4, we noted that $\Delta_{P,F}$ becomes increasingly bad as an approximation for $\Delta_P$ as $d_E$ is increased along the $\beta=5^\circ$ ray. We see this in the cyan, green, and blue curves in the top panel of Figure~1: increasing values of $d_E$ lead to increasing deviation in the final timing perturbation from the value predicted by the plane-wave approximation. Despite this, the inset shows that smaller values of $d_E$ cause the timing perturbation to initially grow more rapidly than what is anticipated by the plane-wave approximation.  

On larger scales than what is shown in Figure~4, it becomes clear that $|\Delta_F/\Delta_{P,F}|$ approaches unity everywhere. When $|\chi|$ and $\psi$ are on the order of 100, $\Delta_{P,F}$ deviates from $\Delta_F$ by approximately one percent.

\section{Near-field BWMs and Pulsar Timing Arrays}

Typical BWM detection scenarios involve several years of pulsar timing preceding the BWM passing through the solar system. This allows the pre-burst rotational parameters\footnote[2]{The observed pulsar periodicity is related to the rest-frame periodicity of a pulsar through approximately constant Doppler shifts.} of the pulsars to be well determined. Then the BWM induces a correlated, simultaneous, and effectively instantaneous change in the apparent rotational frequency of all pulsars. After several additional years of timing, depending on the amplitude of the burst, the orientation of the source relative to the Earth-pulsar systems, and the precision with which the pulsars are timed, the influence of the BWM may be detected. Given the non-linear growth of $\Delta$ in time, derivatives of a pulsar's apparent rotational frequency may also be observed to change and the timing signature will vary from pulsar to pulsar in a manner much different than what is anticipated from very distant BWM sources.

To assess the BWM-induced change in the observed rotational parameters of a pulsar, we model the timing perturbation as a quadratic in time. Figure 1 makes it apparent that in some cases the growth of $\Delta$ cannot be well described by a quadratic, even over only a decade of post-burst timing. However, this model is motivated by a common pulsar timing practice. A pulsar's rotational frequency and its derivative are always included in the timing model that is fit to the observed pulse times of arrival. This has the effect of removing a quadratic from the timing residuals. We write 
\begin{eqnarray}
\label{quadModel}
\frac{\Delta(t)}{\Delta_z}={\cal G}_0t+\frac{1}{2}{\cal G}_1t^2.
\end{eqnarray}
Higher order derivatives are sometimes included in the timing models of young and energetic pulsars, but this is typically not done for the very stable MSPs used in precision PTA efforts as the frequency second derivative is usually too small to be measured. 

\begin{figure}
\begin{center}
\includegraphics[scale=.45]{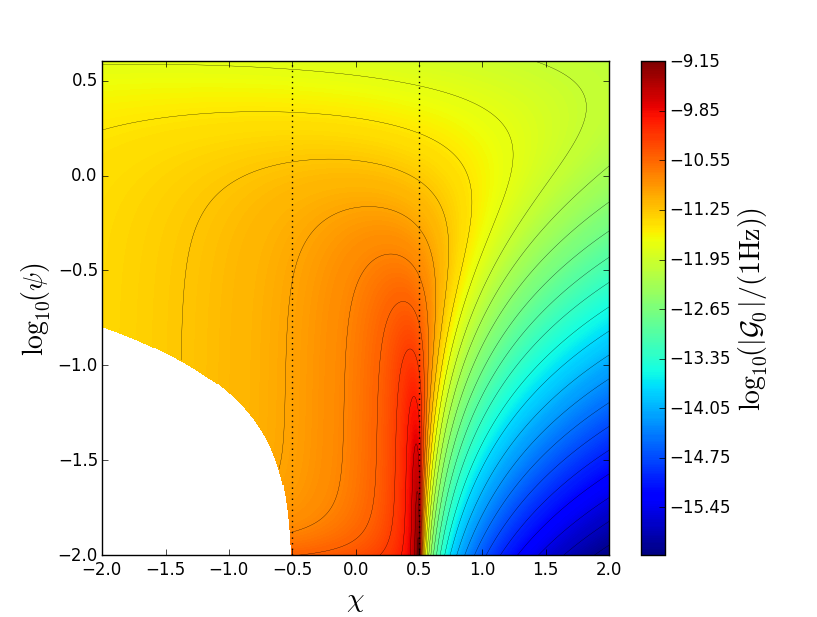}
\includegraphics[scale=.45]{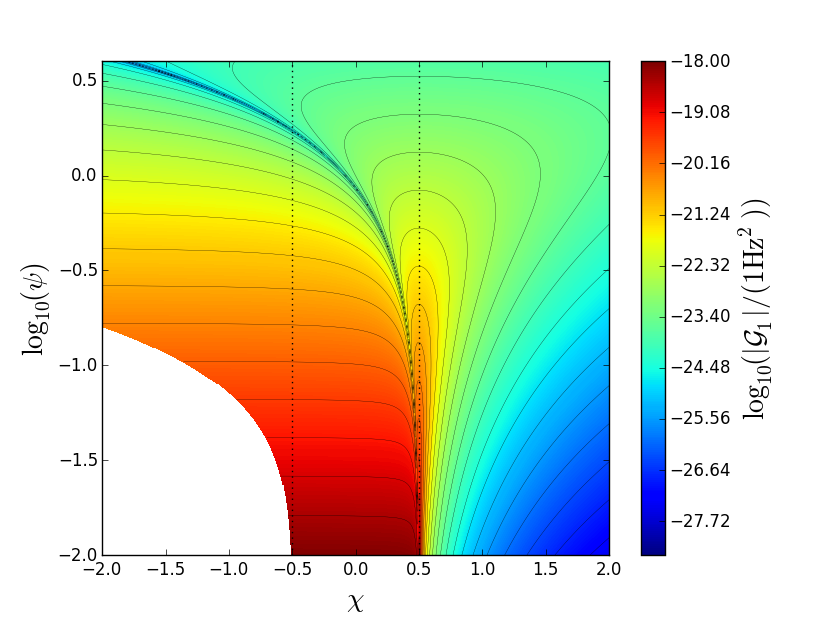}
\includegraphics[scale=.45]{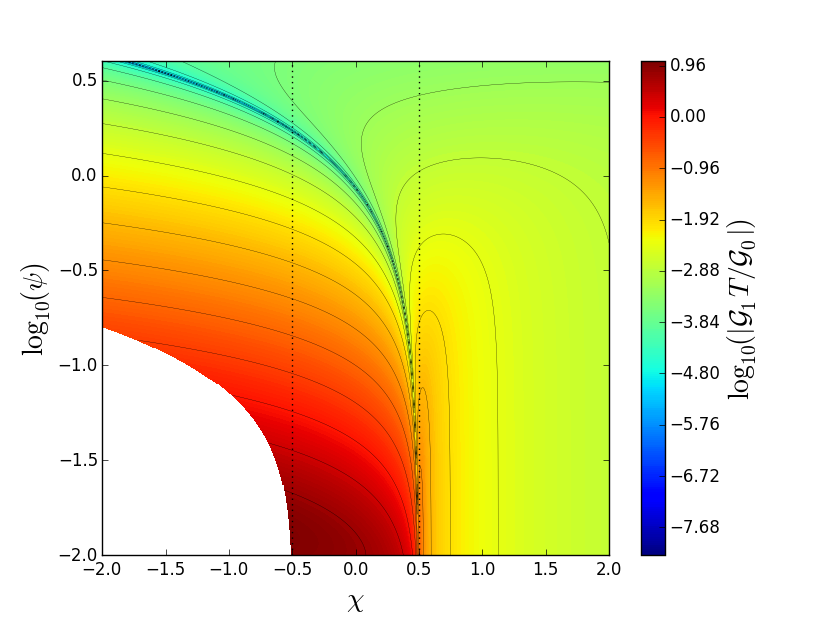}
\caption{Results of fitting a quadratic to $\Delta(t)/\Delta_z$ between $t=0$ and 10 yr. The linear coefficient is ${\cal G}_0$ and the quadratic coefficient is ${\cal G}_1$. The ratio ${\cal G}_1T/{\cal G}_0$ is a measure of the relative importance of the linear versus the quadratic character of the evolution of $\Delta(t)$.}
\end{center}
\end{figure}

Suppose the Earth-pulsar system is again confined to the $x$-$y$ plane (so that we can rescale our results by $\Delta_{\rm SN}$ or $\Delta_m$ and still have correct results for those types of BWM sources) and assume that post-burst, timing data have been accumulated for $T=10$ years. We fit the model described in Equation~\ref{quadModel} to $T$ years of post-burst evolution of $\Delta(t)$. In Figure~5, we show how ${\cal G}_0$ and ${\cal G}_1$ vary in $\chi$ and $\psi$. We also depict the ratio ${\cal G}_1T/{\cal G}_0$ to weigh the relative importance of the linear and quadratic terms over the span of $T$ years. The masked white region in the bottom left corner of each panel indicate regions where $t_F<T$ and $\Delta$ abruptly stops growing within the time $T$; this introduces spurious behavior in the quadratic fitting.

The thin blue ridge seen in the bottom two panels of Figure~5 correspond to the ray of $\beta=60^\circ$ highlighted in Figure~4. For $\beta < 60^\circ$, the quadratic growth of the timing perturbation causes $\Delta$ to grow faster than the linear growth of $\Delta_P$. This is easily seen in the green and cyan curves in the inset of the top panel of Figure~1. For $\beta>60^\circ$, $\Delta$ grows more slowly than is anticipated by $\Delta_P$ due to a change in sign of the quadratic term. For $60^\circ<\beta\lesssim90^\circ$, despite the slower growth of $\Delta$ relative to $\Delta_P$, $\Delta_F$ still manages to exceed $\Delta_{P,F}$ because $\Delta$ continues to grow after $\Delta_P$ has plateaued.

\subsection{Timing Model Fits}

Unless a BWM were accompanied by some other detected event---an electromagnetic or neutrino burst---one would not know if or when the BWM occurred nor that a pulsar's observed rotational properties underwent a sudden change at a certain epoch. Without appropriate modifications to a pulsar's timing model, the timing residuals for that pulsar would begin to show structure unaccounted for by the model. The residuals would not directly show the timing perturbation from the BWM as that signature will be covariant with parameters of the timing model. 

\begin{figure}
\begin{center}
\includegraphics[scale=.41]{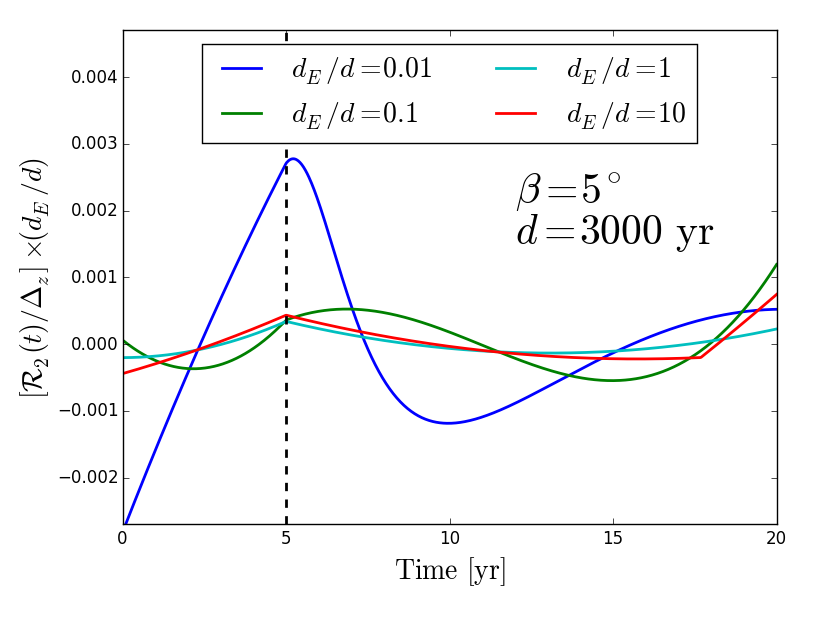}
\includegraphics[scale=.41]{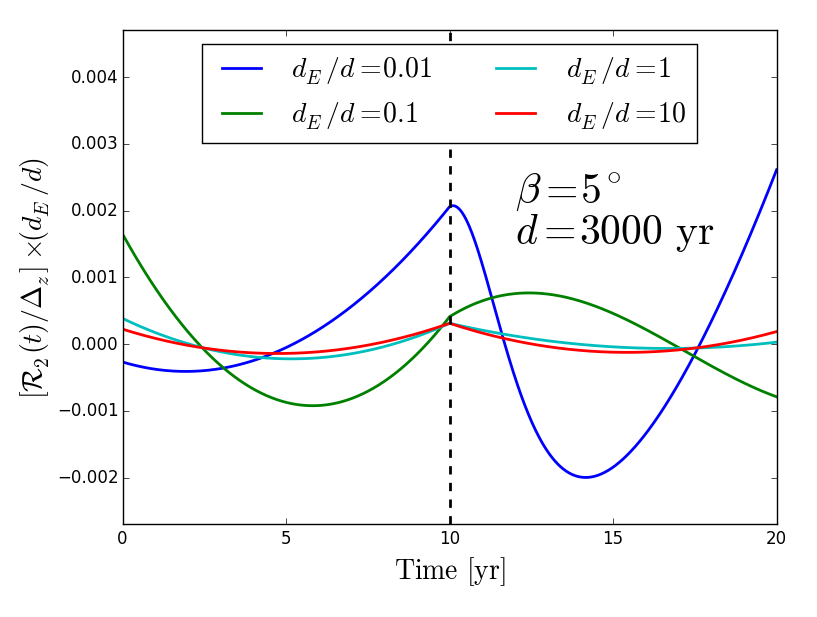}
\caption{The timing perturbation of a BWM with a quadratic removed (${\cal R}_2$ is $\Delta$ with a second-order polynomial removed). The two panels correspond to the BWM occurring at different times within the 20 year window that is shown---5 years in for the top panel and 10 years in for the bottom (the event occurs at the time indicated by the vertical dashed black line). As in the top panel of Figure~1, $\beta=5^\circ$ and the four curves correspond to different values of $d_E$. All curves are multiplied by $d_E/d$ to counteract reductions in the amplitude of the BWM with increases in $d_E$ (Reminder: $d$ is the distance between the Earth and pulsar; $d_E$ is the distance from the BWM source to the Earth). With $d_E/d=10$ (the red curves), the results already begin to closely resemble the result obtained when $d_E/d\rightarrow\infty$.}
\end{center}
\end{figure}

In Figure~6, we show the BWM-induced timing residuals ${\cal R}_2$ with a best-fit quadratic removed as happens when fitting for a pulsar's rotational frequency and its derivative. As with Figure~1, we have multiplied each curve by $d_E/d$ to counteract the fall-off of the wave's amplitude with increasing $d_E$. We show the result when the BWM occurs 5 years (top panel) and 10 years (bottom panel) into a 20 year data set (as indicated by the dashed black line). We have fixed $\beta=5^\circ$ because it is for small $\beta$ that the near-field effects are most pronounced. 

The red and cyan curves of Figure~6, corresponding to the largest source distances, resemble the signature that is anticipated in the plane-wave limit: a double-lobed structure with a kink at the epoch of the BWM. The red curve in the top panel has a second kink later in the 20 year time span. This occurs at $t_F$ when the timing perturbation stops growing (we discuss this in the following subsection). The kinks are essentially indiscernible for the green and blue curves associated with closer sources. Those curves instead resemble smoothly varying cubics or quartics, qualitatively similar to the low frequency timing-noise seen in some pulsars that is  commonly referred to as red spin noise \citep{sc10}.

\subsection{The Pulsar Term}
The moment when a BWM-induced timing perturbation stops growing can introduce a dramatic change point within a pulsar timing span and has been discussed as another opportunity to detect a BWM, potentially thousands of years after the wavefront passed over the Earth \citep{cj12,mcc14}. These so-called ``pulsar term" detections face the problem that the signal stops growing in different pulsars at different times, so a change may be detected in the behavior of an individual pulsar with no contemporaneous detection in other pulsars to validate the behavior as being due to a GW event rather than an anomaly in a particular pulsar (a glitch, for example). 

With the near-field effects of a Galactic BWM, pulsar term detections are plausible because the timing perturbation abruptly stops changing in some instances. We see this prominently in, for instance, the cyan curve in the top panel of Figure~1. We also see the pulsar-term kink in the red curve in the top panel of Figure~6, even after a quadratic has been removed as part of fitting a timing model to the residuals. But this is not always the case. Though there is formally a time $t_F$ when the timing perturbation stops changing, in many instances (e.g. all curves in the top panel of Figure~2) the perturbation gradually approaches $\Delta_F$ and there is no discernible moment when the pulsar term event is observed to occur.

\subsection{Globular Clusters and the Galactic Center}

Conventional PTA efforts to detect GWs seek correlations in the timing residuals of multiple pulsars. The near-field timing effects we have investigated are most pronounced when the BWM source is close to the path of photons traveling to the Earth from a particular pulsar. A single Galactic BWM would produce signatures with markedly different time-domain behavior and widely varied amplitudes for pulsars in different parts of the Galaxy. Globular clusters (GCs) host many MSPs within several light years of each other \cite{rhs+05}. These collections of adjacent pulsars provide opportunities for multiple pulsars to have nearly the same orientation relative to a BWM source, whether the source is Galactic or much more distant. All pulsars in a GC would show similar BWM-induced timing perturbations when the wavefront passes through the solar system and the pulsar term events in those pulsars would all occur within years of each other.

Additionally, the high stellar densities in GCs relative to the Galactic field make them more likely locations for Galactic BWMs to occur. Capture interactions and mergers occur at an enhanced rate in GCs. It is possible that GCs host intermediate mass black holes between $10^2$ and $10^4$ $M_\odot$, but the escape velocity of GCs is on the order of tens of km s$^{-1}$, placing a modest cap on the velocities one can anticipate in a GC \cite{csc13,prf+17,kbl17,psl+17}. 

\begin{figure}
\begin{center}
\hspace*{-0.7cm}
\includegraphics[scale=.43]{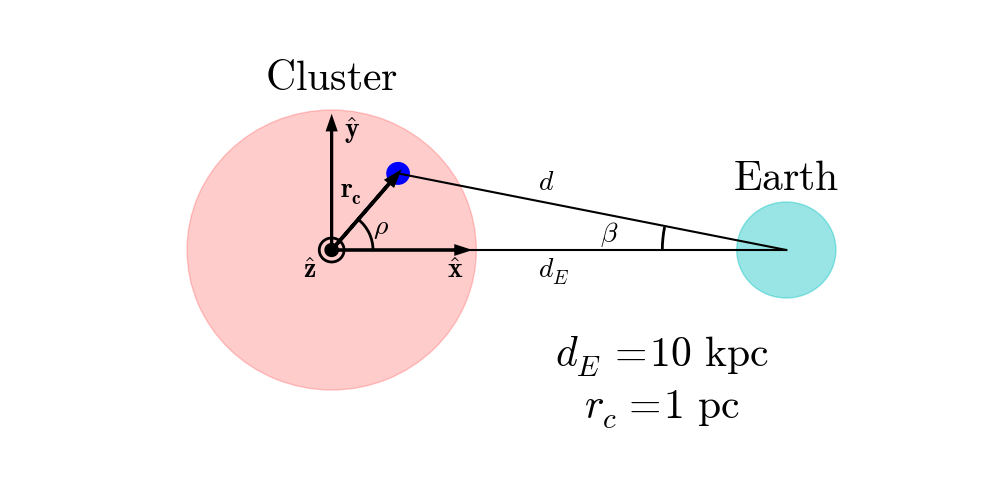}
\includegraphics[scale=.45]{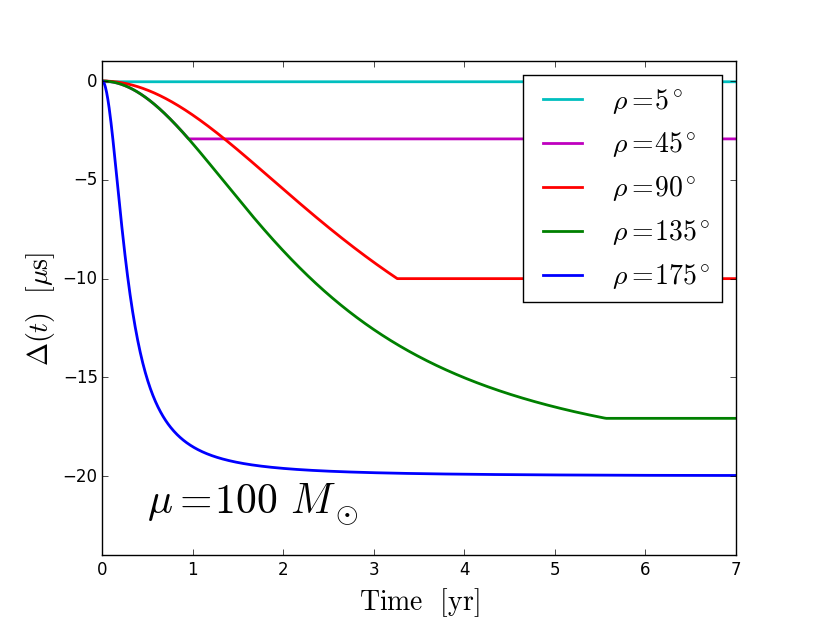}
\caption{{\bf Top}: A schematic representation of a dense cluster, e.g. a globular cluster or the galactic center, as it is oriented relative to the Earth. We consider a BWM source positioned at the center of the cluster and a pulsar at a distance $r_c$ from the cluster's center and at an angle $\rho$ from the line connecting the BWM source and Earth. We consider a cluster 10 kpc from Earth and a pulsar 1 pc from the BWM source. {\bf Bottom}: The timing perturbation produced for a variety of angles $\rho$. We have assumed the BWM source is a merging binary situated in the $x$-$y$ plane, that the Earth and pulsar are also confined to this plane, and that the binary has a reduced mass of 100~$M_\odot$ (setting the magnitude of $\Delta_m$).}
\end{center}
\end{figure}

If a GC were to host a BWM, the impact parameter of the BWM source relative to every cluster pulsar would be on the order of a few lightyears, the characteristic size of the GC. The distance to a GC is on the order of kiloparsecs. To describe the location of a pulsar inside a distant GC, we introduce the parameters $r_c$, the distance of the pulsar from the center of the GC, and $\rho$, the angle between the line connecting the pulsar and GC center and the line connecting the GC center and the Earth (see the top panel of Figure 7). When $r_c\ll d_E$,
\begin{eqnarray}
t_F=\frac{r_c}{c}(1-\cos{\rho})+{\cal O}\left(\frac{r_c^2}{cd_E}\right),
\end{eqnarray}
and if we take the source to be a merging binary with the Earth and pulsar confined to the $x$-$y$ plane,
\begin{eqnarray}
\Delta_F=-\frac{\Delta_m}{2}(1-\cos{\rho})+{\cal O}\left(\frac{r_c}{d_E}\right).
\end{eqnarray}
In less than the light crossing time of the cluster, the timing perturbation reaches $\Delta_F$ which approaches $\Delta_m$ as $\rho$ approaches $\pi$. In the bottom panel of Figure 7, we show $\Delta(t)$ through $t_F$ for 5 different values of $\rho$ assuming $d_E=10$ kpc, $r_c=1$ pc, and $\mu=100~M_\odot$. 

The features of GCs we have discussed are also descriptive of the Galactic center. The high density of stellar and intermediate mass black holes in this region plus the presence of the supermassive black hole Sagittarius A$^*$ make the Galactic center a potential source of a Galactic BWM \cite{krp12}. Additionally, there could be between $10^2$ and $10^3$ pulsars beamed towards Earth within a parsec of the Galactic center \cite{wcc+12,cl14}. Intervening gas and dust strongly disperse and scatter radio pulsations coming from the Galactic center. This causes great challenges for discovering pulsars in this region. But pulsar searches in this region are ongoing and next-generation telescopes and instrumentation will likely begin to reveal this pulsar population \cite{dcl09,bdf+17}.

The timing perturbation from inspiral mergers is proportional to the reduced mass of the system, $\mu$. When the ratio of the more massive object in a binary to that of the less massive object is large, the reduced mass tends towards the mass of the less massive object. So even if there are intermediate mass black holes in GCs with masses of $10^2$ to $10^4~M_\odot$, the reduced mass of any binaries they are in will tend towards the 1 to 10 $M_\odot$ of their probable companion. The presence of Sagittarius~A$^*$ in the Galactic center allows for mergers with significantly higher reduced mass. If a $10^2$ to $10^4$ $M_\odot$ intermediate mass black hole were to spiral into and merge with Sagittarius A$^*$, the pulsars in the vicinity of the Galactic center could show a timing perturbation of tens to thousands of microseconds over just a few years.

\section{Conclusions}

Since pulsars are spread throughtout the Galaxy, it is important to understand how they are influenced by any potential sources of GWs close to them. The simple time domain behavior of BWMs and that they are believed to accompany a wide array of GW events makes them particularly useful for these considerations. The analysis we have done will prove instructive for future efforts to understand other potential Galactic GW sources influencing PTAs.

Because SN occur on the order of once per century per Galaxy, they are among the most reasonable potential sources of a Galactic BWM. We found that given the characteristic energy of a SN, detecting a BWM from such a source through pulsar timing is unlikely. However, rarer events that are more luminous in GWs (and thus memory), such as inspiral mergers of compact remnants, are potentially detectable with fortuitous alignment of the BWM source and well-timed pulsars. Certain astrophysical environments, like globular clusters and the Galactic center, offer an enhanced likelihood of a BWM occurring and an increased concentration of pulsars (all relative to the Galactic field). Though there may be difficulties in precisely timing pulsars in such environments due to gravitational interactions within the dense environments and interstitial gas and dust in the case of the Galactic center, pulsars in these environments may prove useful for detecting Galactic BWMs, particularly if something like a binary black hole merger were to occur. The merger of a binary with a reduced mass on the order of 100 $M_\odot$ could induce a timing perturbation of tens of microseconds over the span of just a few years and the perturbation scales linearly with the reduced mass. This is a large perturbation compared with typical pulsar timing accuracy and could be detected.

As a final thought, near-field effects provide a means by which BWMs can influence pulsar timing efforts not only at the times when the wavefront passes through the solar system and past the pulsar, but also during the entire interregnum between these events, likely thousands of years. For BWMs from extragalactic sources that are well described as plane waves, the timing perturbation grows linearly in time causing the measured rotational period of a pulsar to differ very slightly from its intrinsic value. With no \textit{a priori} knowledge of the pulsar's intrinsic rotational period, this slight deviation will be entirely subsumed into the timing model for that pulsar. 

But near-field effects lead to BWM-induced timing perturbations that evolve in a non-linear, non-polynomial, and sometimes non-monotonic fashion. If SN occur in the Milky Way once or twice per century, the BWMs produced by dozens of SN over the last few millenia, each of which may have been individually undetectable by modern PTAs, could form a non-negligible stochastic GW foreground. Multiple rarer and brighter events, the most energetic classes of SN or mergers of compact objects, may also contribute to this foreground. The properties of such a foreground, both spectral and spatial, and the prospects for its detection by PTAs, will be considered in future work.

\vspace{0.7cm}
DRM is a Jansky Fellow of the National Radio Astronomy Observatory (NRAO). NRAO is a facility of the National Science Foundation (NSF) operated under cooperative agreement by Associated Universities, Inc. This work was supported by the National Science Foundation under Grant No. 1417132 awarded to DFC and through the NANOGrav Physics Frontiers Center award number 1430284.

\appendix
\section{Asymptotic Behavior for Distant BWMs}
First consider $t_F$ as defined in Equation \ref{finalTime}. It can be written as
\begin{eqnarray}
t_F&=&\frac{b}{c}\left[\epsilon+\sqrt{(\cot{\beta}-\epsilon)^2+1}-\sqrt{\cot^2{\beta}+1}\right],
\end{eqnarray}
where $\epsilon=d/b$. Note that $b=d_E\sin{\beta}$. If $\epsilon\ll1$, as it will be in the limit $d_E\rightarrow\infty$, a sound expansion of $t_F$ is
\begin{eqnarray}
t_F&=&\frac{b}{c}\left[\epsilon(1-\cos{\beta})+{\cal O}(\epsilon^2)\right],\nonumber\\
&=&\frac{d}{c}(1-\cos{\beta})+{\cal O}\left(\frac{d^2}{cb}\right).
\end{eqnarray}  
Our solution thus reproduces the lag between $t_0$ and $t_1$ at leading order. The higher order corrections become important for small values of $b$ which can be achieved for even very large values of $d_E$ if $\sin{\beta}$ is sufficiently small. This tension between large values of $d_E$ and small values of $\sin{\beta}$ will appear again in these asymptotic expansions. 

Now we analyze the time evolution of $\Delta$ to show that it asymptotically reduces to $\Delta_P$ for variations of $\varphi$ and $\beta$. For simplicity, we will look at its time derivative since for times between $t_0$ and $t_1$, the time derivative of $\Delta_P$ is a simple constant for fixed values of $\varphi$ and $\beta$:
\begin{eqnarray}
\label{deltaDerivative}
\frac{d\Delta}{dt}&=&-\Delta_z\frac{\alpha_0+\alpha_1u_L+\alpha_2u_L^2}{2(1+u_L^2)^{5/2}}\frac{du_L}{dt}\nonumber\\
&=&-\Delta_z\frac{\alpha_0+\alpha_1u_L+\alpha_2u_L^2}{2(1+u_L^2)^{5/2}}\left(\frac{c}{d_E}\csc{\beta}\frac{du_L}{d\delta}\right),\nonumber\\
\end{eqnarray}
where $\delta\equiv ct/b$. We have already seen based on the behavior of $t_F$ for very large $d_E$ that $t$ grows to, at most, twice the Earth-pulsar light travel time before $u_L\rightarrow u_-$  and the evolution in time halts. Then, for sufficiently large values of $d_E$, again, since $b=d_E\sin{\beta}$, $\delta\ll1$ for all values of $t$.

We consider a specific configuration where the Earth lies in the $x$-$z$ plane, an angle $\theta$ from the positive $z$ axis, i.e. $\hat{\bf d}_E=\sin{\theta}\hat{\bf x}+\cos{\theta}\hat{\bf z}$. Since the memory source we are considering is axisymmetric, there is no loss of generality by confining the Earth to the $x$-$z$ plane. We describe the position of the pulsar relative to the Earth with spherical polar coordinates $\theta_P$ and $\phi_P$:
\begin{eqnarray}
{\bf d}_P-{\bf d}_E=d(\cos{\phi_P}\sin{\theta_P}~&\hat{\bf x}&+\nonumber\\\sin{\phi_P}\sin{\theta_P}~&\hat{\bf y}&+\nonumber\\\cos{\theta_P}~&\hat{\bf z}&). 
\end{eqnarray}
Recognize that $({\bf d}_P-{\bf d}_E)/d=-\hat{\bf k}$. The components of $\hat{\bf b}$ can be found by solving the following system of equations:
\begin{eqnarray}
\hat{\bf b}\cdot(\hat{\bf k}\times\hat{\bf d}_E)&=&0,\\
\hat{\bf b}\cdot\hat{\bf k}&=&0,\\
|\hat{\bf b}|&=&1.
\end{eqnarray}
If $\hat{\bf b}$ is a solution to the above equations, $-\hat{\bf b}$ will be as well. This sign ambiguity is easily resolved by recognizing that in any configuration of Earth, pulsar, and GW source, $\hat{\bf b}\cdot\hat{\bf d}_E=|\sin{\beta}|\geq0$. The solution is
\begin{eqnarray}
\hat{b}_x=b_n^{-1}&&\left[\sin{\theta}(\cos^2{\theta_P}+\sin^2{\phi_P}\sin^2{\theta_P})\right.\nonumber\\
&&\left.-\cos{\theta}\cos{\phi_P}\cos{\theta_P}\sin{\theta_P}\right],\\
\hat{b}_y=b_n^{-1}&&\left[\sin{\phi_P}\sin{\theta_P}(\cos{\theta}\cos{\theta_P}\right.\nonumber\\
&&~~~~~~~~~~~~~~~~~~\left.+\sin{\theta}\cos{\phi_P}\sin{\theta_P})\right],\\
\hat{b}_z=b_n^{-1}&&\left[\sin{\theta_P}(\cos{\theta}\sin{\theta_P}\right.\nonumber\\
&&\left.~~~~~~~~~-\sin{\theta}\cos{\phi_P}\cos{\theta_P})\right],
\end{eqnarray}
where,
\begin{eqnarray}
b_n=&&\left[\sin^2{\theta}\cos^2{\theta_P}\right.\nonumber\\
&&\left.-2\cos{\theta}\sin{\theta}\cos{\phi_P}\cos{\theta_P}\sin{\theta_P}\right.\nonumber\\
&&\left.+\sin^2{\theta_P}(\cos^2{\theta}+\sin^2{\theta}\sin^2{\phi_P})\right]^{1/2}.
\end{eqnarray}

With $\hat{\bf k}$ and $\hat{\bf b}$ fully determined, everything in Equation~\ref{deltaDerivative} is fully determined. If we then expand Equation~\ref{deltaDerivative} in powers of $\delta$, we can express the result as
\begin{eqnarray}
\frac{d\Delta}{dt}=&&\frac{c\Delta_z}{2d_E}\sin^2{\theta}~\cos{\left[2\arctan{\left(\frac{\hat{k}_y}{\hat{\bf k}\cdot(\hat{\bf y}\times\hat{\bf d}_E)}\right)}\right]}\nonumber\\
&&~~~~~~~~~~~~~~\times\left(1+\hat{\bf k}\cdot\hat{\bf d}_E\right)+{\cal O}(\delta).
\end{eqnarray}

To zeroth order in $\delta$, $d\Delta/dt$ is identical to Equation~\ref{planeWaveLimit} if we recognize that  
\begin{eqnarray}
h_B&=&\frac{c\Delta_z}{d_E}\sin^2{\theta},\\
\varphi&=&\arctan{\left(\frac{\hat{k}_y}{\hat{\bf k}\cdot(\hat{\bf y}\times\hat{\bf d}_E)}\right)},\\
\end{eqnarray}
and, again, that $\cos{\beta}=\hat{\bf k}\cdot\hat{\bf d}_E$.

\section{Memory from Supernovae}

In an inertial frame (call it frame B) with coordinates $r$, $\theta$, and $\phi$, let the energy distribution of relativistic particles released by the supernova be 
\begin{eqnarray}
\frac{dE}{d\Omega}=\sum_{l,m}b_{lm}Y_{lm}(\theta,\phi).
\end{eqnarray}
Assume that the particles move radially outward at the speed of light, i.e. ${\bf v}=c\hat{\bf \zeta}$ where $\hat{\bf \zeta}$ points in the direction of an infinitesimal parcel of jettisoned energy. We can express the ${\rm TT}$ metric perturbation from memory a distance $r$ from the origin in direction $\hat{\bf n}$ as
\begin{eqnarray}
\label{appendixStrain}
h^{\rm mem}_{ij,B}=\frac{4G\Theta(t_{\rm ret})}{c^4r}\Lambda_{ij,kl}\int d\Omega \frac{dE}{d\Omega}\frac{\hat{\zeta}^k\hat{\zeta}^l}{1-\hat{\bf n}\cdot\hat{\bf \zeta}}.\nonumber\\
\end{eqnarray} 

The above integral is more readily done in a frame that is rotated relative to frame B (call it frame A with coordinates $\theta'$, $\phi'$, and $r'$) such that $\hat{\bf n}=\hat{\bf z}'$. We will eventually have $\hat{\bf n}$ track a photon along its trajectory, so frame A will rotate dynamically in time relative to frame B. We can re-express the energy distribution in frame A: 
\begin{eqnarray}
\frac{dE}{d\Omega}=\sum_{l,m}a_{lm}Y_{lm}(\theta',\phi').
\end{eqnarray}

Let $R=R_z(\phi_t)R_y(\theta_t)R_z(-\phi_t)$ be the Euler rotation matrix rotating frame B into frame A. $R_y(\theta_t)$ is a rotation by $\theta_t$ about $\hat{\bf y}$ and $R_z$ is similarly defined. The rotation satisfies the condition $R\hat{\bf z}=\hat{\bf n}$. The first rotation about $\hat{\bf z}$ does nothing to $\hat{\bf z}$, but by setting that angle to $-\phi_t$, we ensure that there is no relative rotation between frames A and B when $\theta_t=0$. The coefficients in frame A can be written in terms of the coefficients in frame B as
\begin{eqnarray}
a_{lm}=\sum_{m'}b_{l,m'}e^{i(m'-m)\phi_t}d_{m,m'}^l(\theta_t),
\end{eqnarray} 
where $d_{m,m'}^l$ are Wigner (small) d-matrix elements.

In frame A, the denominator in the integrand of Equation~\ref{appendixStrain} simplifies to $(1-\cos{\theta'})$ and the integration can be carried out: 
\begin{eqnarray}
h_{ij,A}^{\rm mem}=\frac{4G\Theta(t_{\rm ret})}{c^4r}\sum_{l\geq 2}\mu_l(a_{l-2}T_{ij}+a_{l2}T^*_{ij}),
\end{eqnarray}
where
\begin{eqnarray}
T=\left(\begin{array}{ccccc}
1~&&~-i~&&~0\\
-i~&&~-1~&&~0\\
0~&&~0~&&~0\end{array}\right),
\end{eqnarray}
and
\begin{eqnarray}
\mu_l=\sqrt{\frac{(2l+1)\pi}{24\prod_{j=2}^{l-1}(j+3)(j-1)^{-1}}}.
\end{eqnarray}

The metric perturbation in frame A can be converted back to frame B with the Euler rotation $R$. Noting that $R$ is an orthogonal matrix with $R^T=R^{-1}$, 
\begin{eqnarray}
h_{ij,B}^{\rm mem}=R^T_{ik}h_{kl,A}^{\rm mem}R_{lj}.
\end{eqnarray}
Defining $\hat{\bf k}'=R\hat{\bf k}$, 
\begin{eqnarray}
\hat{k}_i\hat{k}_jh^{\rm mem}_{ij,B}=\frac{4G\Theta(t_{\rm ret})}{c^4r}\sum_{l\geq2}\mu_l&&\left[a_{l-2}(\hat{k}'_x-i\hat{k}'_y)^2+\right.\nonumber\\
&&\left.~a_{l2}(\hat{k}'_x+i\hat{k}'_y)^2\right].
\end{eqnarray}

Only moments with $l\geq2$ influence the memory, so the memory from the shell is independent of the total energy or linear momentum of the shell. For illustrative purposes, we will consider a case with $|b_{20}|>0$ but all other moments with $l\geq2$ are zero. In this case,
\begin{eqnarray}
a_{2-2}=a_{22}^*=\sqrt{\frac{3}{8}}b_{20}e^{2i\phi_t}\sin^2{\theta_t},
\end{eqnarray}
and it can be shown that
\begin{eqnarray}
\label{SNmemory}
\frac{\Delta(t)}{\Delta_{\rm SN}}=&&\frac{\sqrt{5\pi}}{2}\int_{u_L(t)}^{u_+}du \frac{\sin^2{\theta_t}}{\sqrt{u^2+1}}\times\nonumber\\
&&[(\hat{k}'_x-\hat{k}'_y)(\hat{k}'_x+\hat{k}'_y)\cos{(2\phi_t)}+2\hat{k}'_x\hat{k}'_y\sin{(2\phi_t)}],\nonumber\\
\end{eqnarray}
where $\Delta_{\rm SN}=Gb_{20}/c^5$

Through $R$, $\hat{\bf k}'$ is a function of $\theta_t$ and $\phi_t$, themselves functions of $u$:
\begin{eqnarray}
\theta_t&=&\arccos{\left(\frac{u\hat{k}_z+\hat{b}_z}{\sqrt{u^2+1}}\right)},\\
\phi_t&=&\arctan{\left(\frac{u\hat{k}_y+\hat{b}_y}{u\hat{k}_x+\hat{b}_x}\right)}.
\end{eqnarray}
To produce Figure~3, we computed the integral in Equation~\ref{SNmemory} numerically.

\bibliographystyle{apsrev4-1}
\bibliography{nearMemoryVI_dfc.bib}
\end{document}